\documentclass{winnower}

\begin{document}

\title{Reductionism and the Universal Calculus}

\author{Gopal P. Sarma\thanks{Email: gopal.sarma@emory.edu}}
\affil{School of Medicine, Emory University, Atlanta, GA, USA}

\date{}

\maketitle

\begin{abstract}
In the seminal essay, ``On the unreasonable effectiveness of mathematics in the physical sciences," physicist Eugene Wigner poses a fundamental philosophical question concerning the relationship between a physical system and our capacity to model its behavior with the symbolic language of mathematics.  In this essay, I examine an ambitious 16th and 17th-century intellectual agenda from the perspective of Wigner's question, namely, what historian Paolo Rossi calls ``the quest to create a universal language.''  While many elite thinkers pursued related ideas, the most inspiring and forceful was Gottfried Leibniz's effort to create a ``universal calculus,'' a pictorial language which would transparently represent the entirety of human knowledge, as well as an associated symbolic calculus with which to model the behavior of physical systems and derive new truths.  I suggest that a deeper understanding of why the efforts of Leibniz and others failed could shed light on Wigner's original question.  I argue that the notion of reductionism is crucial to characterizing the failure of Leibniz's agenda, but that a decisive argument for the why the promises of this effort did not materialize is still lacking.  
\end{abstract}

\section{Introduction}
\setlength{\parindent}{0cm}
By any standard, Leibniz's effort to create a ``universal calculus'' should be considered one of the most ambitious intellectual agendas ever conceived.  Building on his previous successes in developing the infinitesimal calculus, Leibniz aimed to extend the notion of a symbolic calculus to all domains of human thought, from law, to medicine, to biology, to theology.  The ultimate vision was a pictorial language which could be learned by anyone in a matter of weeks and which would transparently represent the factual content of all human knowledge.  This would be the starting point for developing a logical means for manipulating the associated symbolic representation, thus giving rise to the ability to model nature and society, to derive new logic truths, and to eliminate logical contradictions from the foundations of Christian thought \citep{yates, rossi, gomar}. \\

Astonishingly, many elements of this agenda are quite familiar when examined from the perspective of modern computer science.  The starting point for this agenda would be an encyclopedia of structured knowledge, not unlike our own contemporary efforts related to the Semantic Web, Web 2.0, or LinkedData \citep{wolfram}.  Rather than consisting of prose descriptions, this encyclopedia would consist of taxonomies of basic concepts extending across all subjects.  \\

Leibniz then wanted to create a symbolic representation of each of the fundamental concepts in this repository of structured information.  It is the choice of the symbolic representation that is particularly striking.  Unlike the usual mathematical symbols that comprise the differential calculus, Leibniz's effort would rely on \emph{mnemonic images} which were useful for memorizing facts. \\

Whereas modern thinkers usually imagine memorization to be a task accomplished through pure repetition, 16th and 17th-century Europe saw fundamental innovation in the theory and practice of memory.  During this period, practitioners of the memory arts relied on a diverse array of visualization techniques that allowed them to recall massive amounts of information with extraordinary precision.  These memory techniques were part of a broader intellectual culture which viewed memorization as a foundational methodology for structuring knowledge \citep{yates, rossi, gomar}.  \\ 

The basic elements of this methodology were mnemonic techniques.  Not the simple catch phrases that we typically associate with mnemonics, but rather, elaborate visualized scenes or images that represented what was to be remembered.  It is these same memory techniques that are used in modern memory competitions and which allow competitors to perform such superhuman feats as memorizing the order of a deck of cards in under 25 seconds, or thousands of random numbers in an hour.  The basic principle behind these techniques is the same, namely, that a striking and inventive visual image can dramatically aid the memory \citep{foer}. \\

Leibniz and many of his contemporaries had a much more ambitious vision for mnemonics than our modern day competitive memorizers.  They believed that the process of memorization went hand in hand with structuring knowledge, and furthermore, that there were better and worse mnemonics and that the different types of pictorial representations could have different philosophical and scientific implications. \\

For instance, if the purpose was merely to memorize, one might create the most lewd and absurd possible images in order to remember some list of facts.  Indeed, this was recommended by enterprising memory theorists of the day trying to make money by selling pamphlets on how to improve one's memory \citep{yates, gomar}.  To give the reader some intuition for what these techniques look like, let me give a simple example.  Suppose I want to remember that the word feline means ``possessing the qualities of a cat."  I might imagine a huge palace devoted to cats, with two huge statues of cats on either side of the entrance.  I might imagine that a large number of cats were standing in ``line'' waiting to enter the facility, and that the statues were made of ``Fe," or iron.  \\

As absurd and silly as it might seem, mnemonic techniques such as this one are incredibly effective at helping people to remember.  Joshua Foer's memoir \emph{Moonwalking with Einstein} is an engaging and insightful first-person account of the ``competitive memory circuit,'' where techniques such as this one are the bread and butter of how elite competitors are able to perform feats of memory that boggle the mind \citep{foer}. \\

But whereas in the modern world, mnemonic techniques have been relegated to learning vocabulary words and the competitive memory circuit, elite intellectuals several centuries ago had a much more ambitious vision the ultimate implications of this methodology.  In particular, Leibniz hoped that through a rigorous process of \emph{notation engineering} one might be able to preserve the memory-aiding properties of mnemonics while eliminating the inevitable conceptual interference that arises in creating absurdly comical, lewd, or provocative mnemonics.  By drawing inspiration from the Chinese alphabet and Egyptian hieroglyphics, he hoped to create a language that could be learned by anyone in a short period of time and which would transparently-- through the pictorial dimension-- represent the factual content of a curated encyclopedia.  Furthermore, by building upon his successes in developing the infinitesimal calculus, Leibniz hoped that a logical structure would emerge which would allow novel insights to be derived by manipulating the associated symbolic calculus \citep{yates, rossi, gomar}.  \\

\begin{figure}\label{fig:sym-sys}
\begin{center}
\includegraphics[width=.5\textwidth]{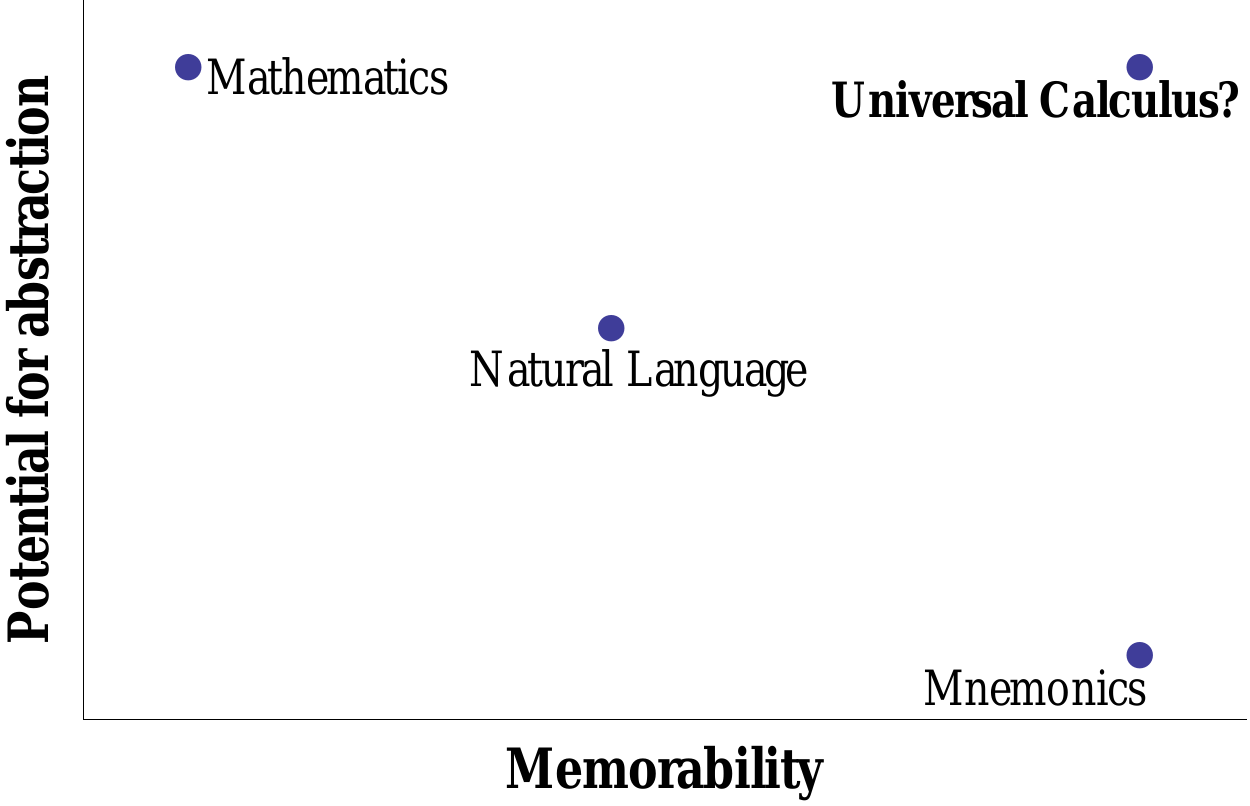}
\caption{{\small Diagram showing where different symbolic systems lie with respect to two criteria central to understanding Leibniz's efforts to create a universal calculus-- memorability and potential for conceptual abstraction.  The point in the upper right indicates a hypothetical system combining the memory-aiding power of mnemonics with the potential for logical abstraction of mathematics. (Figure originally taken from \citep{gomar})}}
\end{center}
\end{figure}

Figure \ref{fig:sym-sys} shows a cartoon diagram which attempts to capture how Leibniz and his contemporaries might have viewed different possible symbolic systems with respect to the capacity for abstraction and the criterion of memorability.  It is the latter capacity which is so incredibly unusual and striking from a modern perspective.  Although mathematical scientists (and also programming language designers) certainly value ease of understanding and notational clarity, the notion that a symbolic language could be lifted off of the page and sealed into one's mind with the certainty of rote memorization is a truly foreign one for contemporary thinkers. \\

Leibniz's motivations extended far beyond the realm of the natural sciences.  Using mnemonics as the core alphabet to engineer a symbolic system with complete notational transparency would mean that all people would be able to learn this language, regardless of their level of education or cultural background.  It would be a truly universal language, one that would unite the world, end religious conflict, and bring about widespread peace and prosperity.  \\

It was a beautiful and humane vision, although it goes without saying that it did not materialize.  

\section{Wigner's question and the failure of Leibniz's agenda}

In a widely cited essay, physicist Eugene Wigner writes:

\begin{quote}
{\small
The miracle of the appropriateness of the language of mathematics for the formulation of the laws of physics is a wonderful gift which we neither understand nor deserve. We should be grateful for it and hope that it will remain valid in future research and that it will extend, for better or for worse, to our pleasure, even though perhaps also to our bafflement, to wide branches of learning \citep{wigner}.
}
\end{quote}

For the remainder of the essay, I will refer to the question ``why does mathematics allow us to formulate physical laws and predict the behavior of Nature?" as ``Wigner's question.''  How does this relate to Leibniz and the universal calculus?  Leibniz believed  that a suitable variant of classical mnemonic techniques would ultimately result in a symbolic system vastly superior to ordinary mathematics.  And coming from a man who had just delivered one of the most important mathematical formalisms ever developed, this was not a vision to scoff at.  Whereas Newton quickly tacked important problems in physics using the tools of the infinitesimal calculus, Leibniz chose instead to view ``ordinary'' calculus as simply a ``toy model'' of a much more ambitious agenda, namely the notion of extending a symbolic calculus to all domains of human knowledge.  \\

In other words, Leibniz viewed mnemonics (or the superior versions of them he hoped to create, drawing inspiration from the Chinese and Egyptian alphabets) as providing a \emph{potential basis for physical theories}.  It is this perspective of Leibniz that provides the link to Wigner's question.  In particular, I suggest the following closely related questions: 

\begin{itemize}
\item Why does mathematics give rise to a near infinite capacity for conceptual abstraction, whereas mnemonics do not?  
\item Why did the agenda of the universal calculus and the vision of a single symbolic language spanning all areas of knowledge fail?  
\end{itemize}
  
As Wigner himself says, most of what will be said about these topics will not be new.  Many, if not most, theoretical scientists have marveled at some point or another about the miracle of mathematics and its relationship to physical systems.  To begin to answer the above questions, I suggest a secondary question which is a converse to Wigner's original one.  The question is the following:  What would a symbolic system look like which had many of the properties of mathematics, but not others?  That is, a symbolic system which comes close to behaving like mathematics, but then falls short, and in particular, which does not allow us to model physical systems and prove its unreasonable effectiveness? \\

My claim is that 16th and 17th-century efforts to create a universal language provide us with an entire category of symbolic systems which are ``non-quite-mathematics.''  If we assume at the most basic level that mathematics is a symbolic language which encodes a set of relationships, then surely mnemonics also have this property.  For instance, the image of the palace of cats that I gave above is a symbolic representation of the word ``feline'' and if I were to make a list of mnemonics which allowed me to memorize the entirety of the periodic table of elements, I would also have a symbolic representation of the corresponding chemical and physical properties.  And yet it seems that there is nowhere to go from here.  Whereas the symbols of mathematics give rise to near infinite capacity for conceptual abstraction, with mnemonics, it seems that we are stuck in the mud.  They start with some mathematical properties in being symbolic representations of certain concepts but they do not build from there in allowing for any conceptual abstraction.  \\

Leibniz had some fascinating ideas for how to build up a logical structure on top of the core system of mnemonics.  For instance, a problem which he viewed as foundational to constructing the universal calculus was to determine for any given predicate, what the valid objects would be, and conversely, given an object, to compute all possible predicates.  One method of attack that he invented was to fix an ordering of predicates, and to each one, assign the corresponding prime number.  Thus, one could uniquely associate to any given object a composite number, from which the prime factorization would give the corresponding predicates, all of which would have associated to them, a mnemonic image.  \\

It is quite an ingenious idea (and reminiscent of G\"odel numbers).  But most importantly, the above example serves to undermine the criticism that the universal calculus failed because it was simply a one-to-one mapping from facts to symbols without an associated logical structure.  Constructing such a structure was precisely Leibniz's aim, although it is not clear exactly how much progress he made \citep{rossi, wolfram}. \\

And the fact that this was even attempted provides us with a novel attack on Wigner's question.  It gives us a set of archeological artifacts in the history of mathematics, symbolic systems which possessed some, but not all of the properties of mathematics, and which ultimately, did not give rise to any capacity to model nature.  In other words, they were quite reasonably ineffective at explaining anything.  Why?  

\section{Reductionism and levels of abstraction}

In this section, I suggest a partial answer to the second question given above.\footnote{I do not believe that the argument I present here addresses the more narrow question of why mnemonics fail at providing an adequate symbolic language capable of giving rise to a rich logical structure and potential for conceptual abstraction.}  My thesis is the following: the failure of Leibniz's agenda can be attributed to an incoherent application of and appreciation for reductionism.  In particular, creating the universal calculus would require symbolic representations in different domains which operated at vastly different levels of abstraction (for example, time scales, or length scales) to be integrated into a single system. \\

Let me now elaborate on the above argument.  Three centuries after Leibniz, we can say purely from experience, that systems amenable to mathematical analysis appear periodically, although it is not entirely clear a priori that this will be the case \citep{wigner, nks}.  Consider, for example, the biological sciences-- which one might have inferred from Rutherford's famous quip-- is a subject doomed to stamp collecting, with the primary work being empirical and the content of the various sub-disciplines largely taxonomic.  And yet in the last century, we have seen an explosion of mathematical work in the biological sciences, ranging from population genetics and evolutionary game theory, to genomics and theoretical neuroscience.  If one had told the mathematical luminaries and physicists of the 18th century that biology would give rise to mathematical models of any level of sophistication, would they have believed it?  Furthermore, the biological sciences are an area where data-driven modeling is giving rise to significant transformation.  And as Halevy, Norvig, and Pereira discuss in their widely read essay, ``The Unreasonable Effectiveness of Data,''  while data-driven methodology is certainly sophisticated in its mathematical machinery, most do not consider this type of modeling to be reductive.  Indeed, their essay is provocatively titled specifically to draw attention to Wigner's question.  The authors suggest that there are many areas where we are unlikely to find fundamental principles and that we should consider phenomenological, data-driven models that have practical value to be genuine intellectual accomplishments worthy of our satisfaction. \citep{norvig}.  \\

To be more precise, let me briefly give examples of what differentiates phenomenological from reductive models-- we will come back to this distinction in more detail in the subsequent section.  A \emph{phenomenological} model refers to a description of a system that captures the directly observed behavior without any reference to the constituent parts of the system.  The etymology of the word refers to the fact that these models simply ``capture the phenomenon''-- i.e. the observed phenomenon-- in spite of the fact that there may very well be more fundamental underlying causal factors. \\

Consider, for example, the use of Punnet squares to model Mendelian inheritance in genetics.  Standard Mendelian genetics was characterized before the discovery of DNA, which is the more fundamental biological mechanism behind the patterns of inheritance that were observed phenotypically.  In other words, by using the simple rules of Mendelian inheritance, we can make predictions-- indeed, quantitative ones-- about the distribution of traits in offspring, in spite of the fact that we are not modeling the behavior of the underlying causal elements \citep{alberts2013essential}. \\

In this case, a phenomenological model is adequate to make precise predictions.  Indeed, a truly reductive model, where we modeled all of the interactions of DNA replication, mitosis, and meiosis, etc. would be far too complicated and unnecessary.  Yet, it is not always clear when a phenomenological model is adequate, or when a reductive model is necessary but computationally intractable. \\

As a counterpoint, consider the growth of mathematical techniques in quantitative finance.  At the most basic level, the aim of financial models is to make predictions of financial time series from historical data.  Like Mendelian genetics and Punnet squares, these models are phenomenological.  That is, we do not try to predict the behavior of financial time series by modeling the mental states of every person involved in the stock market and related financial institutions.  Unlike the case of genetics, these models have yet to show substantial predictive power, despite their considerable mathematical sophistication \citep{derman2004my}.  \\

On the other hand, a reductive model is one that captures the interaction between the most fundamental constituents of a system, such as atoms or fundamental particles.  Some of the most successful theories in physics, such as quantum mechanics or general relativity, are considered to be reductive, although as I explore in the next section there is not a rigorous understanding of what differentiates a reductive from phenomenological model in many cases.\footnote{There is considerable philosophical literature examining the validity of reductionism as a general scientific principle.  One classic argument \emph{against} reductionism is Putnam and Fodor's \emph{multiple realizability argument} \citep{putnam1980nature, putnam1980philosophy, fodor1968psychological, fodor1975language}.  See also Sober's criticism of these arguments \citep{sober1999multiple}. \\

The aim of this article is somewhat distinct from that of philosophers examining reductionism critically as a general scientific principle.  In particular, while my proposed connection between Leibniz' agenda and Wigner's question is indeed a philosophical one, Leibniz's agenda itself was quite a practical one, and the question of why this agenda failed, or could not have worked might just as easily be cast as a question about the failure of an engineering design.  Therefore, my choice of the word ``reductionism'' in this context-- specifically, as denoting a key failure point in Leibniz' agenda-- is used in the sense of practical scientific modeling and not in the sense of its potential philosophical implications more broadly conceived.  \\

As an example, consider that scientific modelers who are simulating the behavior of a system may be forced to choose between a fully reductive model and a phenomenological one purely because of computational intractability.  This is a practical consideration and not a philosophical one-- the availability of faster computers might make the fully reductive model tractable at some future point (see, for example, the discussion in the subsequent section on the use of adiabatic elimination techniques in quantum optics).  \\

Nonetheless, there is certainly overlap between these two agendas.  Although I do not explore this connection further, let me briefly mention work from the philosophy of science community that may be of interest to some readers.  In a recent article, Eleanor Knox proposes a line of thought which may be relevant to the current investigation \citep{knox2016abstraction}.  In particular, she argues that the change from levels of abstraction can introduce \emph{novel explanation}, and furthermore that a theory can possess novel explanatory power even when a reductive theory is available.  Although I leave Knox's analysis of abstraction, reductionism, and emergence unexplored in this manuscript, I mention it here for its possible relevance to my proposed diagnosis of the failure of Leibniz' agenda.  Specifically, Knox's proposal that abstraction can give rise to novel explanatory power relative to a more reductive theory suggests that Leibniz' calculus-- were it to have materialized-- would likely consist of incompatible elements as a result of the mixing of phenomenological and reductive models.  I mention this connection for completeness and leave further analysis for future investigation.  
}  \\

The fundamental observation to take away from these examples is that it is not at all clear what it would mean for all of the diverse theories and models spanning all human endeavors-- whether phenomenological models or reductive ones-- to be integrated into a single symbolic system.  They exist at different levels of abstraction, meaning that the fundamental constituents of each theory can vary widely from one theory to the next.  In quantum mechanics, the fundamental objects are wave functions in Hilbert space.  In general relativity, they are solutions to Einstein's field equations.  In biology, they might be idealized genomes represented by a linear sequence of base pairs, whereas in theoretical neuroscience, they might be phenomenological models of a neuron incorporating many different approximations.  Even setting aside the role of mnemonic techniques in creating this calculus, what could it possibly mean for all of these different theories to exist alongside each other in a way that would allow for a consistent symbolic representation and associated calculus?

\subsection{Universal calculus and computer algebra}
If we are generous, there is one clear analogue to Leibniz's vision in the contemporary world, and that is a computer algebra system (CAS), such as \emph{Mathematica}, Maple, or Maxsyma.  A computer algebra system is a piece of software which allows one to perform specific mathematical computations, such as differentiation or integration, as well as basic programming tasks, such as list manipulation, and so on.  As the development of CAS's have progressed hand in hand with the growth of the software industry and scientific computing, they have come to incorporate a large amount of functionality, spanning many different scientific and technical domains \citep{cohen, mbook}.  \\

In this sense, CAS's have some resemblance to Leibniz vision.  \emph{Mathematica}, for example, has all of the special functions of mathematical physics as well as data from many different sources which can be systematically and uniformly manipulated by the symbolic representation of the Wolfram Language.  One could reasonably claim that it incorporates many of the basic desiderata of Leibniz's universal calculus-- it has both the structured data of Leibniz's hypothetical encyclopedia, as well as symbolic means for manipulating this data. \citep{mbook, wolfram}  \\

However, Leibniz's vision was significantly more ambitious than any contemporary CAS can make claim to have realized.  For instance, while a CAS incorporates mathematical knowledge from different domains, these domains are effectively different modules within the software that can be used in a standalone fashion.  Consider, for example, that one can use a CAS to perform calculations relevant to both quantum mechanics and general relativity.  The existence of both of these capabilities in a single piece of software says nothing about the long-standing theoretical obstacles to creating a unified theory of quantum gravity.  Indeed, as has long been bemoaned in the formal verification and theorem proving communities, CAS's are effectively a large number of small pieces of software neatly packaged into a single bundle that the user interacts with in a monolithic way.  This fact has consequences for those interested in the robustness of the underlying computations, but in the present context, it simply serves to highlight a fundamental problem in Leibniz's agenda \citep{ballarin}.\footnote{To be fair to Leibniz, he may have been aware of some of these conceptual issues-- see the next section on phenomenological models.}  \\

So in effect, one way to describe Leibniz's universal calculus, was an attempt to create something like a modern computer algebra system, but which extended across all areas of human knowledge.  This goal itself would be quite an ambitious one, but in addition Leibniz wanted the additional property that the symbolic representation should have a transparent relationship to the corresponding encyclopedia, as well as possess the capacity of mnemonics to be memorized with ease.  To quote Leibniz himself,
\begin{quote}
{\small
My invention contains all the functions of reason: it is a judge for controversies; an interpreter of notions; a scale for weighing probabilities; a compass which guides us through the ocean of experience; an inventory of things; a table of thoughts; a microscope for scrutinizing things close at hand; an innocent magic; a non-chimerical cabala; a writing which everyone can read in his own language; and finally a language which can be learnt in a few weeks, traveling swiftly across the world, carrying the true religion with it, wherever it goes. \citep{Leibniz}
}
\end{quote}

It difficult to not be swept away by the beauty of Leibniz's imagery.  And yet, from our modern vantage point, there is hardly a doubt that this agenda could not possibly have worked.  

\subsection{Phenomenological models}

There is some reason to believe that Leibniz was quite aware of the issues with mixing phenomenological and reductive models.  Indeed, it appears that an ongoing dialogue between Descartes and Leibniz took place with regards to this exact issue.  The discussion related to the notion of ``true philosophy,'' which I will interpret to mean an adequately reductive model.\footnote{I suspect that a proper investigation into the vision and controversies surrounding the universal calculus would require a thorough understanding of what precisely Leibniz and Descartes meant by this phrase.  To the best of my knowledge, an appropriate modern translation of ``true philosophy'' is an adequately reductive model, but there may be important subtleties that this translation ignores, and perhaps, more importantly, subtleties that this translation adds that could not have possibly existed at the time.}  According to Rossi, while Descartes believed that a universal language was theoretically possible, it would require the development of a true philosophy, which he saw as a fundamental obstacle.  Rossi writes,

\begin{quote}
{\small In a letter to Mersenne in November 1629 \ldots Descartes clearly outlined what he saw as the principal characteristics and aims of a philosophical language, but ultimately his statements are rather ambiguous.  He obviously considered the idea of a philosophical language to be a theoretical possibility.  If one could `establish an order for all the thoughts of the human spirit, in the same way that there exists a naturally established order of numbers,' he suggested, a `language' composed of easily and quickly understood characters could be constructed.  The invention of this language would depend, however, on `the prior construction of a true philosophy, because otherwise it would be impossible to enumerate all the thoughts of men and put them into their correct order.' \citep{rossi}
}
\end{quote}

To this criticism, Leibniz responded:

\begin{quote}
{\small Although this language (the universal calculus) depends on true philosophy, it does not depend on its perfection.  Let me just say this: this language can be constructed despite the fact that philosophy is not perfect.  The language will develop as scientific knowledge develops.  While we are waiting, it will be a miraculous aid: to help us understand what we already know, and to describe what we do not know, and help us to find the means to obtain it, but above all it will help us to eliminate and extinguish the controversial arguments which depend on reasons, because once we have realized this language, calculating and reasoning will be the same thing. \citep{couturat}
}
\end{quote} 

In modern theoretical science, the distinction between phenomenological and reductive modeling is often a heuristic one and varies wildly from one subject to another.  For instance, in theoretical particle physics, phenomenology refers to efforts to predict and model specific experiments taking place at particle accelerators, or making small extensions to the Standard Model, whereas ``theory'' refers to more fundamental efforts to develop beyond the Standard Model formulations.  And yet, given the wildly successful nature of quantum field theory and the Standard Model, what is considered to be phenomenology in the particle physics community is quite reductive when viewed in comparison to other branches of science that are substantially less mathematically developed.  And there are even rarer cases, such as in theoretical quantum optics, where the existence of rigorous mathematical results from probability theory and functional analysis allows for an interpretation of ``phenomenological model'' that is quite particular to this subject.\footnote{There are several uses of the term ``phenomenological'' in the quantum optics community.  One common usage, which is in line with its use in other communities, is in situations where additional Hamiltonian terms are added to a model to incorporate interactions that are not easily described exactly.  For instance, see Carmichael's classic book \citep{carmichael} for a phenomenological model of atomic dephasing in a radiatively damped two-level atom in thermal equilibrium with a quantized field.  \\

Some recent mathematical developments have allowed for the term ``phenomenological'' to be used in a context that is quite specific to quantum optics.  Here, I am referring to what is known as the Bouten-van Handel-Silberfarb adiabatic elimination theory \citep{Bout08}, a rigorous set of mathematical techniques for constructing reduced order quantum optical models for systems where there are two physical processes of widely separated time scales.  If one is only interested in the dynamics of the system with respect to the slow degrees of freedom, an effective description that eliminates the fast degrees of freedom can simplify calculations and decrease the computational requirements for simulating the system \citep{Kerc10, Kerc11}. \\

In theoretical quantum optics, adiabatic elimination techniques have been used at a heuristic level for several decades and developing an intuition for applying these methods is part of the bread and butter toolkit of traditional quantum optics (see for example, \citep{stockton, gardiner}).  In recent years, a powerful set of tools has emerged that allow for practical physical modeling to be carried out with a formalism that is a precise, non-commutative generalization of classical stochastic calculus \citep{filtering}.  In this framework, the Heisenberg equations of motion correspond to a quantum stochastic differential equation (QSDE), which describes the joint evolution of a system (such as an atom) in interaction with a dense spectrum of free Bosonic fields (such as a laser).  \\

The mathematical formalism of quantum stochastic differential equations has allowed the heuristic techniques of adiabatic elimination to be put on rigorous mathematical footing.  In particular, the idea of two processes of widely differing time scales can be framed in the context of a limit theorem in which a sequence of quantum stochastic models is constructed which depend on a parameter.  The Bouten-van Handel-Silberfarb theory gives a precise set of technical conditions under which the sequence of parameterized equations converges to a limiting QSDE in which the fast degrees of freedom have been eliminated.  \\

In the present context, the rigorous adiabatic elimination techniques serve to demonstrate the point that the distinction between phenomenological and reductive models is quite an informal one and can vary widely from subject to subject.  In particular, because of the availability of the more recent limit theorems for adiabatic elimination, reduced order models that were simply inferred on the basis of physical intuition are often called ``phenomenological`` by quantum opticians who are firmly in the mathematical physics camp.  Note, however, that what distinguishes a phenomenological from a reductive model in this context is not the model itself, but the procedure by which one arrived at the model!  If one simply guessed it on the basis of physical intuition, it would be considered (by some) to be phenomenological because one did not strictly apply a more fundamental technique.  \\

I mention this anecdote from quantum optics simply to make the point that there is no rigorous understanding of what is meant by phenomenological versus reductive models and that in a subject such as quantum optics, where the theory is highly refined, one can make quite subtle distinctions that do not even apply to other branches of physics.  This observation is just one more piece of evidence suggesting that Leibniz's project was bound to fail.  Surely, neither Leibniz nor any other thinker of that time period could have anticipated the rich, surprising, and heuristic nature in which mathematics often turns up across the different branches of science.}\\

So while the conceptual problems with mixing phenomenological and reductive models seem very much to have been part of the broader dialogue among those contemporaries of Leibniz attempting to engineer a universal language, it seems that they could not have possibly appreciated the scope of the issues involved.  Science was likely too young and there were simply not enough examples of mathematical models for Leibniz and his contemporaries to truly appreciate what they were trying to achieve. \\

And yet it is not clear that our own contemporary understanding is much more sophisticated.  While many subjects have a rich intuition honed from decades and centuries of trial and error, such as which approximations are realistic or when simulations can be trusted over analytical models etc., there seems to be little if any deeper understanding of why and how mathematics shows up in the profound and unexpected ways that it does.  In other words, it hardly seems that we understand the unreasonable effectiveness of mathematics any more than Leibniz did.  \\

In declaring Leibniz' agenda to be a failure, am I claiming that the agenda itself was not feasible-- even in principle-- or that Leibniz and his contemporaries did not have the knowledge to execute it?  I believe that we can only give a partial answer to this question, and it is partly a matter of perspective.  If one views efforts in modern computer science as being realizations of Leibniz' agenda, then we can say that Leibniz and the society that he inhabited simply did not have the knowledge to execute this vision.  On the other hand, an integral component of the universal calculus was the role of mnemonics, which have no place whatsoever in modern theoretical science.  The idea that mnemonics themselves could form the basis for a symbolic system, and that this might allow one to create a pictorial language that could be memorized with ease and which would transparently represent the factual content of a curated encyclopedia remains a bit of an enigma.  At present, I do not believe that the arguments I have presented above with regards to reductionism speak to this issue.  That being said, one wonders if it would not be possible for carefully constructed experiments in cognitive science to demonstrate the limitations on using mnemonics as the basis for a symbolic system.  

\section{Conclusion}

In this essay, I have tried to provide a novel perspective by framing a question that is a converse to Wigner's original question.  That is, rather than ask, ``why does mathematics allow us to model natural phenomena?'' the question I have asked is, ``16th and 17th-century efforts to create a universal language provide us with a set of symbolic systems that do not seem to give rise to the conceptual abstraction of mathematics, and yet there was a fierce conviction that these systems would someday supersede mathematics.  Why did this agenda fail?'' \\

Whether one views Leibniz's project as a decisive failure, or a profound set of ideas centuries ahead of their time-- in light of the surprising commonalities with modern efforts at data curation, knowledge representation, and computer algebra-- one concept clearly emerges when trying to characterize the fundamental obstacles to the creation of a universal calculus.  That concept is reductionism.  As I have argued above, the rich interplay between reductive and phenomenological modeling is a fundamental aspect of modern theoretical science, and it is difficult to imagine that Leibniz could have fully anticipated these developments and the consequences for his vision.  Still, it does not seem that our own fundamental understanding is all that sophisticated, although we do have the benefit of a rich cultural intuition that draws on centuries of trial and error. \\

It is difficult to imagine that anything resembling a conclusive ``answer'' to Wigner's question will be formulated without profound philosophical breakthroughs that are unlikely to be anticipated.  Still, like many, I believe that it is a worthwhile question to keep in the back of our minds as we do practical day-to-day work.

\acknowledgments 
\subsection*{Acknowledgements}
I would like to thank Michael Rosen, Rob Spekkens, and Kritika Yegnashankaran for insightful discussions and critical reading of the manuscript.  

\bibliographystyle{abbrvnat}
\bibliography{reductionism}

\end{document}